\documentclass[aps,prl,reprint,groupedaddress,superscriptaddress]{revtex4-1}

\usepackage{graphicx}
\usepackage{dcolumn}
\usepackage{bm}
\usepackage{amsmath}
\usepackage{amssymb}
\usepackage{subfigure}
\usepackage{xcolor}

\begin{document}


\title{Electro-osmotic Instability of Concentration Enrichment in Curved Geometries for an Aqueous Electrolyte}
\author{Bingrui Xu}
\thanks{contributed equally}
\author{Zhibo Gu}
\thanks{contributed equally}
\affiliation{Department of Aeronautics and Astronautics, Fudan University, Shanghai, 200433, China}
\author{Wei Liu}
\affiliation{School of Aerospace Engineering and Applied Mechanics, Tongji University, Shanghai, 200092, China}
\author{Peng Huo}
\affiliation{Department of Aeronautics and Astronautics, Fudan University, Shanghai, 200433, China}
\author{Yueting Zhou}
\affiliation{School of Aerospace Engineering and Applied Mechanics, Tongji University, Shanghai, 200092, China}
\author{S. M. Rubinstein}
\affiliation{John A. Paulson School of Engineering and Applied Sciences, Harvard University, Cambridge, Massachusetts 02138, USA}
\author{M. Z. Bazant}
\affiliation{Department of Chemical Engineering and Department of Mathematics, Massachusetts Institute of Technology, Cambridge, Massachusetts 02139, USA}
\author{B. Zaltzman}
\author{I. Rubinstein}
\affiliation{Department of Mathematics and the BIDR, Ben-Gurion University of the Negev, Sede Boqer Campus, 8499000, Israel}
\author{Daosheng Deng}
\email{dsdeng@fudan.edu.cn}
\affiliation{Department of Aeronautics and Astronautics, Fudan University, Shanghai, 200433, China}
\date{\today}

\begin{abstract}
We report that an electro-osmotic instability of concentration enrichment in curved geometries for an aqueous electrolyte, as opposed to the well-known one, is initiated exclusively at the enriched interface (anode), rather than at the depleted one (cathode). For this instability, the limitation of unrealistically high material Peclet number in planar geometry is eliminated by the strong electric field arising from the line charge singularity. In a model setup of concentric circular electrodes, we show by stability analysis, numerical simulation, and experimental visualization that instability occurs at the inner anode, below a critical radius of curvature. The stability criterion is also formulated in terms of a critical electric field and extended to arbitrary (2d) geometries by conformal mapping. This discovery suggests that transport may be enhanced in processes limited by salt enrichment, such as reverse osmosis, by triggering this instability with needle-like electrodes.


\end{abstract}

\maketitle

\emph{Introduction.}---Ion transport through an aqueous electrolyte plays an essential role in numerous electrochemical technologies, such as fuel cells, flow batteries, electrodialysis and capacitive deionization for water desalination \cite{Probsteinbook, RevNature}. Under applied DC current or voltage, due to the electromigration and diffusion, ions are redistributed spatially to produce concentration polarization \emph{i.e.}, the regions of concentration depletion and enrichment. As the concentration near an ion-selective interface (the membrane, electrode, or nanochannel) becomes strongly depleted by diffusion limitation, the current as a function of voltage tends to saturate at the limiting value, followed by the region of over-limiting conductance.

Several mechanisms for this have been discussed \cite{nikonenko2014,Dukin1991}. In bulk liquid electrolytes, the main physical mechanism for over-limiting current likely is non-equilibrium electro-osmotic instability (NE-EOI)  \cite{rubinstein2000electroosmotically, zaltzman2007electroosmotic, HanPRL2007, rubinstein2008direct, Yossifon2008}.  NE-EOI relies on the transformation of the electric double layer on the electrode or membrane causing salt depletion from its typical nanoscale quasi-equilibrium structure to a non-equilibrium structure of extended space charge \cite{rubinstein2000electroosmotically, zaltzman2007electroosmotic}.  During NE-EOI, over-limiting current is sustained by vortical flows at a macroscopic scale (~100 $\mu$m) in the depleted region, which have been shown to result from instability in parallel-plate (1d) geometries, or in a threshold-less manner in systems with a broken symmetry \cite{HanPRL2007, rubinstein2008direct, Yossifon2008}.
In confined systems, new transport mechanisms can arise. In particular, over-limiting current can be sustained by surface conduction or electro-osmotic flow along the confining surfaces in microchannels \cite{Dydek2011overlimiting,yaroshchuk2011,nielsen2014,KimPRL2015} or porous media \cite{deng2013overlimiting,alizadeh2019,deng2015water,schlumpberger2015,conforti2020}.

Recently, equilibrium electro-osmotic instability  has been invoked  as a possible mechanism of the overlimiting current by relaxing the perfect charge selectivity or infinite conductivity assumptions \cite{RubinsteinEqui2015, RubinsteinPRF2016}. In principle, equilibrium electrokinetic instability in a perfect plane-parallel electrochemical cell has been ruled out, since the required minimal material Peclet number ($\mathrm{Pe_{min}} = 8$) is about one order of magnitude higher than that of the typical aqueous electrolyte solution ($\mathrm{Pe_{aqu}}\simeq 0.5$) \cite{RubinsteinPF1991, Zholkovskij1996}. Physically, for quasi-equilibrium electric double layer, the limiting electroosmotic slip velocity is proportional to the tangential concentration gradient \cite{zaltzman2007electroosmotic}. In the concentration depletion region near the interface (such as the cathode), a seeding vortex is suppressed by a negative feedback, since the descending portion of vortex towards interface brings the high bulk concentration to the cathode. In the concentration enrichment region, the opposite is true, making the instability possible, if only the material Peclet number were an order of magnitude larger.

In this paper, we demonstrate that equilibrium electrokinetic instability can indeed appear exclusively in the region of \emph{concentration enrichment} for an aqueous solution, by eliminating the limitation of unrealistically high Pe in planar geometry via the strong electric field due to the singularity of line charges. This electro-osmotic instability of concentration enrichment (EOI-CE) is different from all the previously studied non-equilibrium or equilibrium EOI occurring in the depletion region.  We predict EOI-CE by numerical simulation and by extending stability analysis of Zholkovskiij,\emph{ et al.} ~\cite{Zholkovskij1996} for the prototypical case of concentric circular electrodes, and the stability criterion is extended to arbitrary 2d geometries by conformal mapping.  Experiments are performed for an aqueous CuSO$_4$ solution in a circular copper electrodeposition cell, confirming this instability.


\begin{figure}[t]
\includegraphics[width=\linewidth]{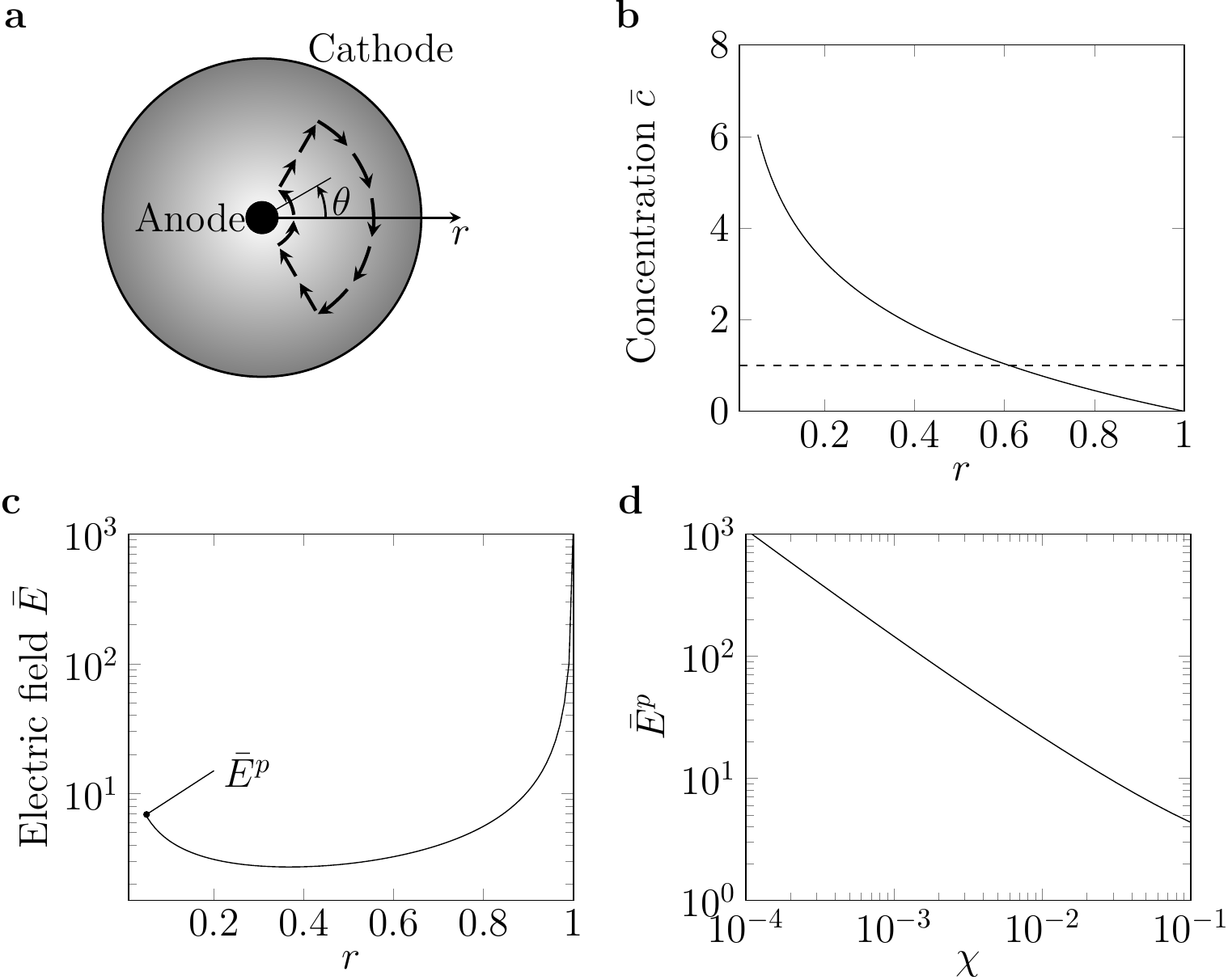}
\caption{ (a) Sketch of the anode vortex near the inner anode in a circular channel. At limiting current, both the concentration enrichment (b) and an additional peak ($\bar{E}^p$) of electric field (c) appear near the inner anode. (d) $\bar{E}^p$ as a function of $\chi$ in a log-log scale, demonstrating the tendency of developing a singularity as $\chi\rightarrow 0$.
\label{fig:sketch}
}
\end{figure}

\emph{Mathematical model.}-- The prototypical model for EOI-CE is a dilute, binary $z : z$ electrolyte of concentration ($c_0$) in an annular channel with an inner anode of radius $R_1$ and outer cathode of radius $R_2$  (Figure \ref{fig:sketch}a).  Geometrical curvature is controlled by the ratio $\chi=R_1/R_2<1$.  For ion transport, there are two regions, the outer bulk electroneutral region valid in the segment $\chi < r < 1$ and the inner electric double layer region valid in the $\varepsilon$-vicinity of the interfaces , where $\varepsilon=(dRT/4z^2F^2\pi c_0)^{1/2}/R_2$ is the dimensionless Debye length \cite{zaltzman2007electroosmotic}. In the bulk electroneutral region, the governing equations are the dimensionless Nernst-Planck-Stokes equations \cite{zaltzman2007electroosmotic}:
\begin{subequations}
		\label{goveqs}
		\begin{align}
		\bar{c}_t+\mathrm{Pe} (\boldsymbol{v} \boldsymbol{\cdot} \nabla )\bar{c} &= \nabla \boldsymbol{\cdot} (\nabla \bar{c} + \bar{c} \nabla \bar{\varphi}),\label{catnp}\\
		\bar{c}_t+\mathrm{Pe} (\boldsymbol{v} \cdot \nabla )\bar{c} &=\nabla \boldsymbol{\cdot} (\nabla \bar{c} - \bar{c} \nabla \bar{\varphi}), \label{annp}\\
          -\nabla \bar{p} +\Delta \bar{\boldsymbol{v}} &=\frac{1}{\mathrm{Sc}}\bar{\boldsymbol{v}}_t, \label{momeq}\\
		\nabla \boldsymbol{\cdot}\bar{\boldsymbol{v}}=0, & \quad \bar{\boldsymbol{v}}=\bar{u} \boldsymbol{i}_r+\bar{w}\boldsymbol{i}_{\theta},
		\end{align}
	\end{subequations}
where $\bar{c}=c^{+}=c^{-}$ is the ionic concentration scaled by $c_0$, $\bar{\varphi}$ is the electric potential scaled by the thermal voltage, $RT/zF$, and $\bar{\boldsymbol{v}}$ is the fluid velocity scaled by $(RT/zF)^2(d/4\pi\eta R_2)$, where $d$ is the dielectric constant and $\eta$ the dynamic viscosity of the solution; for simplicity, the ionic diffusivities are assumed to be equal, $D_{+}=D_{-}=D$; the solution is governed by the material Peclet number $\mathrm{Pe}=(RT/zF)^2(d/4\pi\eta D)$ and the Schmidt number, $\mathrm{Sc}=\nu/D$, since position is scaled to the outer radius $R_2$ and time to the diffusion time, $R_2^2/D$.

The dimensionless boundary conditions at each interface are
	\begin{subequations}
	\label{bcons}
	\begin{align}
		\left( \bar{c}_r-\bar{c} \varphi_r\right) \lvert_{r=\chi, 1}=0, &\quad \ln \bar{c}+\bar{\varphi}\lvert_{r=\chi}=\ln p_1+V,\\
		\ln \bar{c}+\bar{\varphi}\lvert_{r=1}=\ln p_1, &\quad \bar{u}\lvert_{r=\chi, 1}=0,\quad \bar{w}\lvert_{r=\chi, 1}=w_s,
	\end{align}
	\end{subequations}
where $p_1$ is the fixed charge in the cation-selective interface, which could be a perm-selective membrane or electrode (as in our experiments below) and  $w_s$ is the equilibrium electro-osmotic slip from the electric double layer.  Hereafter we assume $V>0$ for the inner anode case to explore the electro-osmotic instability of concentration enrichment.

\begin{figure}[b]
\includegraphics[width=0.8\linewidth]{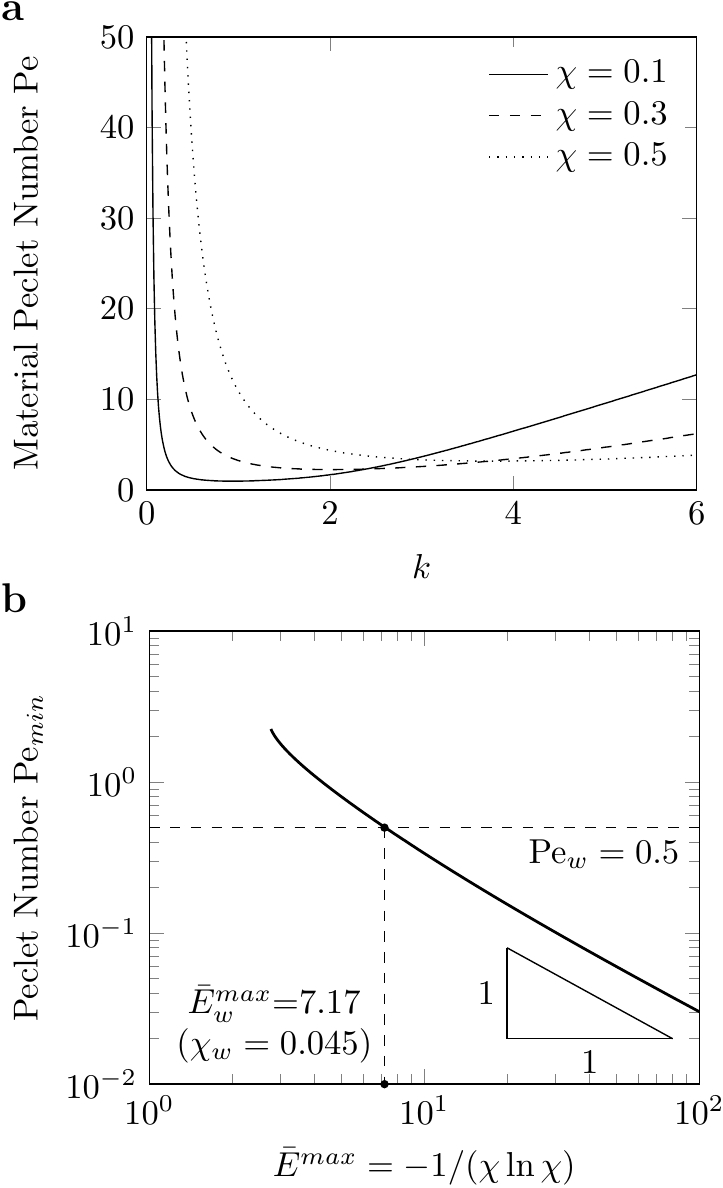}
\caption{Linear stability analysis. (a) The neutral stability curves versus dimensionless wavenumber $k$ for various radius ratios $\chi$, showing  the critical material Peclet number $\mathrm{Pe_{min}}$ for each curve. (b) Nearly inverse dependence of $\mathrm{Pe_{min}}$ on the maximum electric field on the anode, $\bar{E}^{max}=-(\chi \ln \chi)^{-1}$. For aqueous electrolyte with $\mathrm{Pe_{aqu}} = 0.5$, instability occurs for $\bar{E}^{max}>\bar{E}^{max}_{w}=7.17$ or $\chi<\chi_{w}=0.045$.
\label{fig:linearstability}
}
\end{figure}

\begin{figure*}[t]
\includegraphics[width=\linewidth]{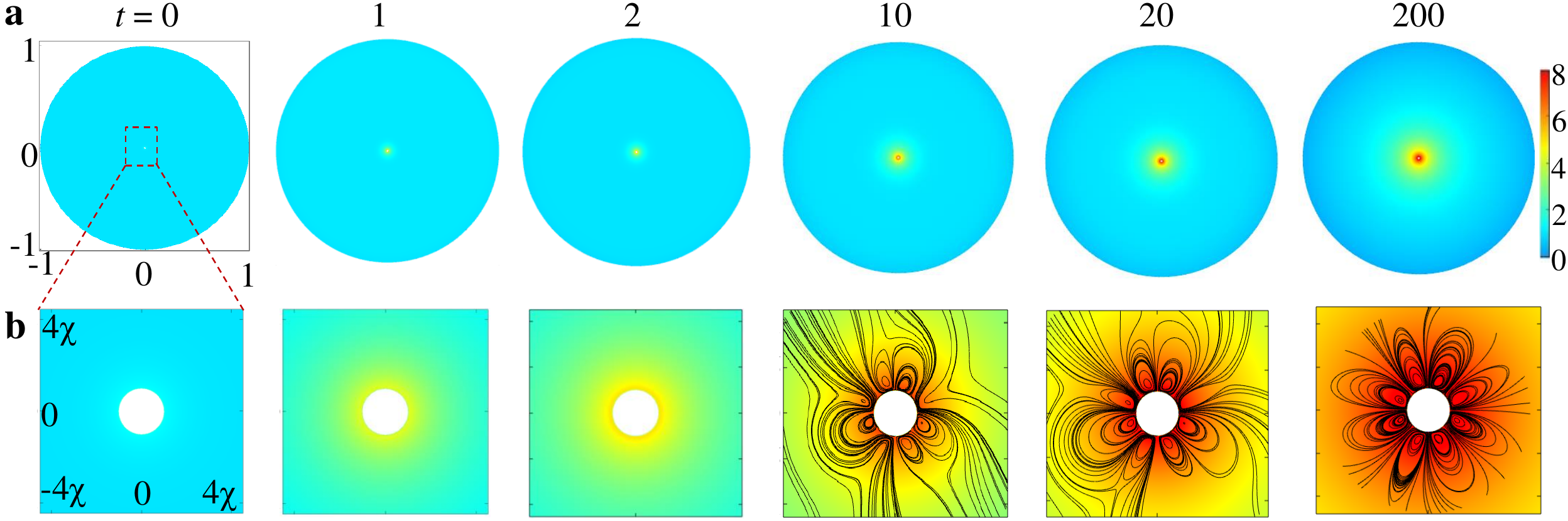}
\caption{Numerical simulation. (a) Top panel for the concentration evolution. (b) Bottom panel for the magnified view around the inner anode, demonstrating the development of EOI and the formation of vortex indicated by the black streamlines in the concentration enrichment region. ($\chi=1/120$, $V=8$, $\varepsilon = 3\times 10^{-4}$, $p_1=2$, and time scaled by $\chi^2R_2^2/D$.)
\label{fig:DNS}
}
\end{figure*}

\emph{One-dimensional quiescent solution.}---The complicated equations \eqref{goveqs} and \eqref{bcons}  admit an exact solution for the quiescent state of steady conduction with $\bar{\boldsymbol{v}} = 0$ in any two-dimensional geometry~\cite{bazant2004}.
For the circular geometry, the outer bulk electroneutral equations can be integrated to obtain the ionic fluxes \cite{zaltzman2007electroosmotic, Dydek2011overlimiting, Gu2019}:
\begin{equation}
	\label{NPcationaion}
	\frac{\mathrm{d}\bar{c}}{\mathrm{d}r}+\bar{c}\frac{\mathrm{d}\bar{\varphi}}{\mathrm{d}r}=-\frac{I}{2\pi r}, \quad \frac{\mathrm{d}\bar{c}}{\mathrm{d}r}-\bar{c}\frac{\mathrm{d}\bar{\varphi}}{\mathrm{d}r}=0,
\end{equation}
where $I$ is the cation flux (equal to current density), scaled to the diffusion limited current.  The solution of Eq.~\eqref{NPcationaion} is
\begin{equation}
\label{concentrationcurrent}
\bar{c}(r)=1-\frac{I}{4\pi}(\ln r+\frac{1}{2}+\frac{\chi^2\ln \chi}{1-\chi^2}).
\end{equation}
Then the electric field is given by
\begin{equation}
\bar{E}=-\frac{\mathrm{d}(\mathrm{\ln}\bar{c})}{\mathrm{d}r} \to -\frac{1}{r \mathrm{\ln}r}, \ \ \mbox{ as } I\to 1.
\end{equation}
According to Eq. \eqref{concentrationcurrent}, the ion concentration is enriched near the inner anode and depleted near the outer cathode (Figure \ref{fig:sketch}b). However, in contrast to the parallel geometry \cite{Zholkovskij1996}, the electric field has two peaks (Figure \ref{fig:sketch}c). One peak at $r=1$ is related to concentration depletion near the outer cathode, common in the concentration polarization phenomenon. The other at $r=\chi$  of magnitude, $\bar{E}^{\max}(\chi)=-1/\chi \mathrm{\ln}\chi$, is a special feature of the circular channel (Figure \ref{fig:sketch}d), which becomes singular in the limit of vanishing anode radius, $\chi\rightarrow 0$, and provides the driving force for instability.

\emph{Linear stability analysis.}---To quantitatively investigate this possible electro-osmotic instability of concentration enrichment by identifying the instability growth rate ($\omega$) dependent on the perturbation wavenumber ($k$), we perform the linear stability analysis of equations \eqref{goveqs} and \eqref{bcons} at the limiting current, extending the results of Ref. \cite{Zholkovskij1996} to curved interfaces [Supplementary Materials (SM)]. In a circular channel, the equilibrium electro-osmotic slip velocity for the inner electric double layer valid in the $\varepsilon$-vicinity of the interfaces is $w_s=\left[4\ln 2 \frac{1}{r}\frac{\partial (\ln \bar{c}) }{\partial \theta}  \right ] \rvert_{r=\chi, 1}$  at $r = \chi, 1$ \cite{Zholkovskij1996} (SM).

The neutral stability curves for growth rate $\omega=0$ (Figure \ref{fig:linearstability}a) indicate a minimum material Pelect number $\mathrm{Pe_{min}}$ for various $\chi$, above which the EOI-CE is possible. This $\mathrm{Pe_{min}}(\chi)$ decreases dramatically with $\chi$ (Figure \ref{fig:linearstability}b), and once  $\chi<\chi_{\mathrm{cri}}$, $\mathrm{Pe_{min}} < \mathrm{Pe_{aqu}}$ ($\mathrm{Pe_{aqu}}\simeq 0.5$ is the typical material Peclet number of an aqueous electrolyte solution \cite{Zholkovskij1996}).

Therefore, the electro-osmotic instability of concentration enrichment is possible in a circular channel ($\chi< \chi_{\mathrm{cri}}$). Furthermore, the growth rate is always negative for $\boldsymbol{v}\lvert_{r=\chi}=0, \boldsymbol{v}\lvert_{r=1}=w_s$, implying no instability at the outer cathode, but the instability can occur at the inner anode. In contrast to the planar-parallel case \cite{Zholkovskij1996}, the dimensionless inner radius $\chi$ provides another degree of freedom to control $\mathrm{Pe_{min}}$, allowing its reduction by more than an order of magnitude, so that $\mathrm{Pe_{min}}<\mathrm{Pe_{aqu}}$ for $\chi <\chi_{\mathrm{cri}}$. This opens the possibility of instability, even for an aqueous electrolyte.

For equal ionic diffusivities and perfectly perm-selective interface with infinite lateral conductivity, assumed in this paper, the electrical body force alone cannot yield instability for material Peclet number relevant for low molecular electrolytes \cite{baygents1998electrohydrodynamic,aleksandrov2002numerical,story2007bulk}. Indeed, this force is negligible from scaling analysis, and has little effect on the neutral stability curves by stability analysis (SM).

\begin{table*}
	\caption{Comparison between EOI-CE and NE-EOI in the aqueous electrolyte solution (and SM).}
	\begin{ruledtabular}
		\begin{tabular}{cccccccc}
			Type&Geometry & Current & Concentration & Mechanism &Size & Velocity& Vorticity \\
			\hline
			NE-EOI & Plane-parallel or circular & Overlimiting  & Depletion region & Extended space charge & 500 $\mu$m & 10 $\mu$m/s & 0.1 s$^{-1}$\\
			EOI-CE & Circular with an inner anode & Limiting  & Enrichment region & Enhanced electric field & 200 $\mu$m & 0.5 $\mu$m/s & 0.001 s$^{-1}$\\
		\end{tabular}
	\end{ruledtabular}
\label{tab:comp}
\end{table*}

\begin{figure*}[t]
\includegraphics[width=\linewidth]{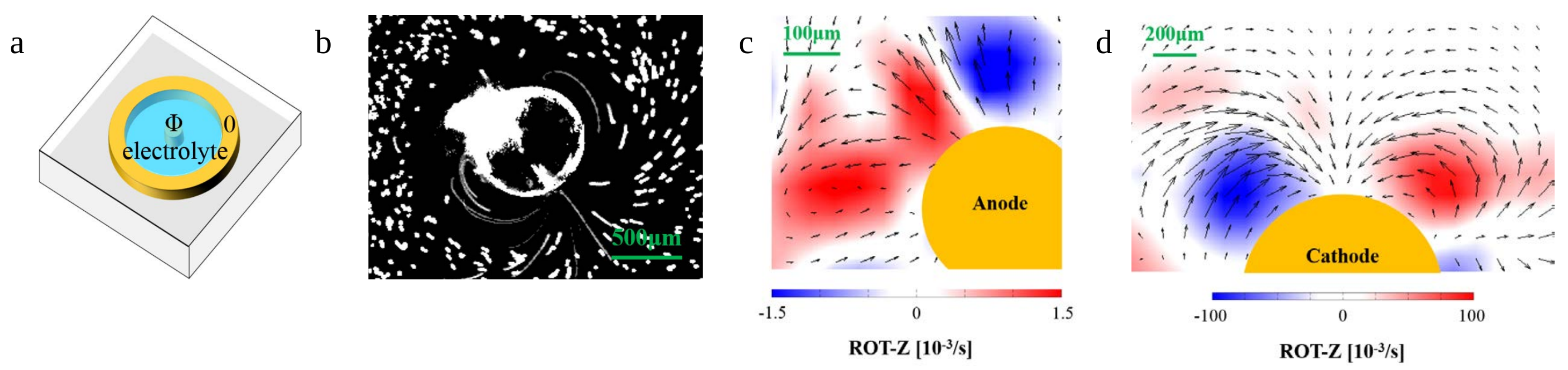}
\caption{Experimental observation of anode vortex in a circular channel. (a) The sketch of the electrochemical cell for a circular channel. (b) The observation of the anode vortex by the time-lapse imaging indicating the streamline ($\chi = 1/60, I = 5 \; \mu A$). PIV imaging of the flow field (c) near the inner anode ($\chi = 1/120, \Phi = 6 \; V$) and (d) near the inner cathode ($\chi = 1/120, \Phi = -6 \; V$).
\label{fig:expobs}
}
\end{figure*}

\emph{Generalization by conformal mapping.}-- Linear stability of EOI-CE is dominated by the local physics of equilibrium electro-osmotic convection near the anode surface, so the stability criterion, expressed in terms of the local electric field, should hold for any locally smooth geometry, if the surface is not too highly curved. In particular, the most unstable wavelength should be smaller than the radius of curvature.  For this regime in the circular geometry ($1/k \ll \chi$), the linear stability criterion plotted in Fig.~\ref{fig:linearstability}(b) is well approximated by the inverse relation, $\mathrm{Pe_{min}}/\mathrm{Pe_{aqu}} < \bar{E}^{max}_w/\bar{E}^{max}$, where $\mathrm{Pe_{aqu}}=0.5$ and $\bar{E}^{max}_w=7.17$.

This suggests a general stability criterion based on a bound for the dimensionless local electric field on the anode in the quiescent state,
\begin{equation}
\mbox{ Stable if:} \ \ \bar{E}^{max} < \alpha\//\mathrm{Pe},
\end{equation}
where $\alpha\approx 14.34$.  Using the general solution of the Nernst-Planck equations in two dimensions~\cite{bazant2004,bazant2003}, $\bar{E}^{max}$ can be derived for any anode geometry (SM):
\begin{equation}
\bar{E}^{max} = \mbox{max}\left\{ \frac{ |f^\prime(z)| \, I }{1 + I },\, \mbox{Im} f(z)=1 \right\},
\end{equation}
where $w=f(z)$ is the conformal map from the electrolyte domain in the complex $z$-plane to the strip, $|\mbox{Im}\, w|<1$, representing parallel-plate electrodes in the mathematical $w$-plane.
 For the circular geometry (Fig. \ref{fig:sketch}), which is defined by the conformal map, $f(z) = -i(1- 2\log z / \ln\chi)$, we recover the result above, $\bar{E}^{max}=-1/(\chi\ln\chi)$, at the limiting current, $I=1$.

\emph{Two-dimensional numerical simulation.}---In order to explore the nonlinear evolution of the instability, we perform numerical simulation of the coupled Nernst-Planck-Poisson and Navier-Stokes equations \cite{Maniphyfluid2013, ManiJCIS2015, LiuJAP2018,LiuPRE2020} (SM and Supplementary Video 1) for the prototypical circular geometry. The concentration evolution (Figure \ref{fig:DNS}a) shows the development of concentration depletion near the outer cathode and concentration enrichment near the inner anode ($\chi=1/120<\chi_\mathrm{cri}$). From Eq. \eqref{concentrationcurrent}, $c(\chi)\approx -2 \mathrm{ln}\chi \approx 10$ in the steady state is comparable with the concentration enrichment near the anode in the simulation. The magnified view of the concentration enrichment region around the inner anode (Figure \ref{fig:DNS}b) demonstrates the development of EOI and the formation of vortex indicated by the black streamlines. Moreover, at a large ratio such as $\chi=1/10>\chi_\mathrm{cri}$, the vortex disappears in the simulation, hence for EOI-CE a sufficient smaller $\chi$ is necessary, which is consistent with the linear stability analysis.

\emph{Experimental observation.}---The electro-osmotic instability of concentration enrichment in the prototypical circular geometry is further verified by experiments. A circular channel embedded in a PDMS device (Figure \ref{fig:expobs}a) is designed \cite{Gu2019}, H $\approx$ 40 $\mu m$ for the channel height, 2$R_2$ = 24 mm  for the outer copper ring, and 2$R_1$ = 2.4, 0.4 and 0.2 mm for the inner copper wire, corresponding to $\chi =1/10, 1/60$ and $1/120$. The aqueous CuSO$_4$ solution is 1 mM, and $\Phi=6$ V (a Keithley 2450 Source Meter) for the inner anode. As expected, for $\chi =1/10 >\chi_{cri}$, no instability is found in the region of concentration enrichment near the central anode. However, for $\chi <\chi_{cri}$, as predicted by the stability analysis and numerical simulations, the streamlines from 20-seconds time-lapse imaging (Figure \ref{fig:expobs}b, $\chi =1/60$) (fluorescent microscope, Zeiss, Axio Zoom V16) \cite{Gu2019} are clearly observed near the central anode, evidently demonstrating the existence of the EOI-CE. Furthermore, by employing particle image velocity (Figure \ref{fig:expobs}c, $\chi =1/120$), this anode vortex is week with velocity about 1 $\mu \mathrm{m/s}$.

\emph{Discussion.}--- We also compared our described  electro-osmotic instability of concentration enrichment with the NE-EOI for the inner cathode (Figure \ref{fig:expobs}d) in the circular channel via experiments (Table \ref{tab:comp}) and simulations (SM). Experimentally, the anode vortex is much weaker with its vorticity about two orders of magnitude lower than the cathodic one, and its velocity about 1 $\mu$m/sec or lower. Consequently, as opposed to the cathode vortex in the concentration depletion region enhancing the ion transport and sustaining the overlimiting conductance, the current is not affected much by the presence of anode vortex which is  confirmed in simulations (SM).

As noted in the Introduction for a plane-parallel geometry \cite{Zholkovskij1996}, theoretically the equilibrium EOI instability is feasible though only for unrealistically high material Peclet number (Pe$>\mathrm{Pe_{min}}$). Indeed, the simulations in this planar geometry (Supplementary Video 2 for Pe = 50, V = 4) confirm that the EOI occurs in the concentration enrichment region near the anode side, not the concentration depleted near the cathode side. Furthermore, this positive (negative) feedback near the concentration enrichment (depletion) interface is clearly unraveled by an illustration of the concentration and potential perturbation from a seeding vortex upon a basic quiescent concentration polarization steady state (SM).

The prerequisite for this equilibrium electro-osmotic instability is exclusively in the enriched region, \emph{i.e.}, as opposed to that addressed by Rubinstein and Zaltzman \cite{zaltzman2007electroosmotic,RubinsteinEqui2015}, this instability is initiated at the enriched interface (anode), rather than at the depleted one (cathode). Also the high electric field at the inner anode due to line charge singularity eliminates the limitation of unrealistically high Pe observed by Zholkovsky \emph{et al.} for planar geometry \cite{Zholkovskij1996}, hence allowing the experimental observation of this instability in an aqueous solution here.

This reported instability has practical implications to enhance membrane processes, such as reverse osmosis, by convective mixing of the concentrated brine layer, stimulated by needle-like curved electrodes driving EOI-CE. For example, cylindrical hollow membrane fiber is a central element of desalination by reverse osmosis \cite{Greenleea2009Reverse} and energy conversion by pressure retarded osmosis \cite{Achilli2009Power} employing salinity variation, while salt accumulation at the membrane is the major limitation of these processes. Hence, introducing the anode into the fiber and reducing the salt accumulation near the anode fiber through instability could be a major remedy for this limitation.


\begin{thebibliography}{22}
\makeatletter
\providecommand \@ifxundefined [1]{%
 \@ifx{#1\undefined}
}%
\providecommand \@ifnum [1]{%
 \ifnum #1\expandafter \@firstoftwo
 \else \expandafter \@secondoftwo
 \fi
}%
\providecommand \@ifx [1]{%
 \ifx #1\expandafter \@firstoftwo
 \else \expandafter \@secondoftwo
 \fi
}%
\providecommand \natexlab [1]{#1}%
\providecommand \enquote  [1]{``#1''}%
\providecommand \bibnamefont  [1]{#1}%
\providecommand \bibfnamefont [1]{#1}%
\providecommand \citenamefont [1]{#1}%
\providecommand \href@noop [0]{\@secondoftwo}%
\providecommand \href [0]{\begingroup \@sanitize@url \@href}%
\providecommand \@href[1]{\@@startlink{#1}\@@href}%
\providecommand \@@href[1]{\endgroup#1\@@endlink}%
\providecommand \@sanitize@url [0]{\catcode `\\12\catcode `\$12\catcode
  `\&12\catcode `\#12\catcode `\^12\catcode `\_12\catcode `\%12\relax}%
\providecommand \@@startlink[1]{}%
\providecommand \@@endlink[0]{}%
\providecommand \url  [0]{\begingroup\@sanitize@url \@url }%
\providecommand \@url [1]{\endgroup\@href {#1}{\urlprefix }}%
\providecommand \urlprefix  [0]{URL }%
\providecommand \Eprint [0]{\href }%
\providecommand \doibase [0]{http://dx.doi.org/}%
\providecommand \selectlanguage [0]{\@gobble}%
\providecommand \bibinfo  [0]{\@secondoftwo}%
\providecommand \bibfield  [0]{\@secondoftwo}%
\providecommand \translation [1]{[#1]}%
\providecommand \BibitemOpen [0]{}%
\providecommand \bibitemStop [0]{}%
\providecommand \bibitemNoStop [0]{.\EOS\space}%
\providecommand \EOS [0]{\spacefactor3000\relax}%
\providecommand \BibitemShut  [1]{\csname bibitem#1\endcsname}%
\let\auto@bib@innerbib\@empty
\bibitem [{\citenamefont {Probstein}(2003)}]{Probsteinbook}%
  \BibitemOpen
  \bibfield  {author} {\bibinfo {author} {\bibfnamefont {R.~F.}\ \bibnamefont
  {Probstein}},\ }\href@noop {} {\emph {\bibinfo {title} {Physicochemical
  Hydrodynamics}}}\ (\bibinfo  {publisher} {Wiley, New York},\ \bibinfo {year}
  {2003})\BibitemShut {NoStop}%
\bibitem{RevNature} M. A. Shannon, P. W. Bohn, M. Elimelech, J. G. Georgiadis,
B. J. Marinas, A. M. Mayes,  Nature, \textbf{452}, 301, (2018).
\bibitem{nikonenko2014} V. V. Nikonenko, A. V. Kovalenko, M. K. Urtenov,
N. D.  Pismenskaya, J. Han, P. Sistat, G. Pourcelly, Desalination \textbf{342}, 85 (2014).
\bibitem{Dukin1991} S. S. Dukhin, Adv. Coll. Interf. Sci. \textbf{35}, 173 (1991).
\bibitem [{\citenamefont {Rubinstein}\ and\ \citenamefont
  {Zaltzman}(2000)}]{rubinstein2000electroosmotically}%
  \BibitemOpen
  \bibfield  {author} {\bibinfo {author} {\bibfnamefont {I.}~\bibnamefont
  {Rubinstein}}\ and\ \bibinfo {author} {\bibfnamefont {B.}~\bibnamefont
  {Zaltzman}},\ }\href@noop {} {\bibfield  {journal} {\bibinfo  {journal}
  {Physical Review E}\ }\textbf {\bibinfo {volume} {62}},\ \bibinfo {pages}
  {2238} (\bibinfo {year} {2000})}\BibitemShut {NoStop}%
\bibitem [{\citenamefont {Zaltzman}\ and\ \citenamefont
  {Rubinstein}(2007)}]{zaltzman2007electroosmotic}%
  \BibitemOpen
  \bibfield  {author} {\bibinfo {author} {\bibfnamefont {B.}~\bibnamefont
  {Zaltzman}}\ and\ \bibinfo {author} {\bibfnamefont {I.}~\bibnamefont
  {Rubinstein}},\ }\href@noop {} {\bibfield  {journal} {\bibinfo  {journal}
  {Journal of Fluid Mechanics}\ }\textbf {\bibinfo {volume} {579}},\ \bibinfo
  {pages} {173} (\bibinfo {year} {2007})}\BibitemShut {NoStop}%
\bibitem{HanPRL2007} S. J. Kim, Y. C. Wang, J. H. Lee, H. Jang, and J. Han, \textbf{99}, 044501 (2007).
\bibitem [{\citenamefont {Rubinstein}\ \emph {et~al.}(2008)\citenamefont
  {Rubinstein}, \citenamefont {Manukyan}, \citenamefont {Staicu}, \citenamefont
  {Rubinstein}, \citenamefont {Zaltzman}, \citenamefont {Lammertink},
  \citenamefont {Mugele},\ and\ \citenamefont
  {Wessling}}]{rubinstein2008direct}%
  \BibitemOpen
  \bibfield  {author} {\bibinfo {author} {\bibfnamefont {S.~M.}\ \bibnamefont
  {Rubinstein}}, \bibinfo {author} {\bibfnamefont {G.}~\bibnamefont
  {Manukyan}}, \bibinfo {author} {\bibfnamefont {A.~D.}\ \bibnamefont
  {Staicu}}, \bibinfo {author} {\bibfnamefont {I.}~\bibnamefont {Rubinstein}},
  \bibinfo {author} {\bibfnamefont {B.}~\bibnamefont {Zaltzman}}, \bibinfo
  {author} {\bibfnamefont {R.~G.~H.}\ \bibnamefont {Lammertink}}, \bibinfo
  {author} {\bibfnamefont {F.~G.}\ \bibnamefont {Mugele}}, \ and\ \bibinfo
  {author} {\bibfnamefont {M.}~\bibnamefont {Wessling}},\ }\href@noop {}
  {\bibfield  {journal} {\bibinfo  {journal} {Physical Review Letters}\
  }\textbf {\bibinfo {volume} {101}},\ \bibinfo {pages} {236101} (\bibinfo
  {year} {2008})}\BibitemShut {NoStop}%
\bibitem{Yossifon2008} G. Yossifon and H. C. Chang, Physical Review Letters \textbf{101,} 254501 (2008).
\bibitem [{\citenamefont {Dydek}\ \emph {et~al.}(2011)\citenamefont {Dydek},
  \citenamefont {Zaltzman}, \citenamefont {Rubinstein}, \citenamefont {Deng},
  \citenamefont {Mani},\ and\ \citenamefont {Bazant}}]{Dydek2011overlimiting}%
  \BibitemOpen
  \bibfield  {author} {\bibinfo {author} {\bibfnamefont {E.~V.}\ \bibnamefont
  {Dydek}}, \bibinfo {author} {\bibfnamefont {B.}~\bibnamefont {Zaltzman}},
  \bibinfo {author} {\bibfnamefont {I.}~\bibnamefont {Rubinstein}}, \bibinfo
  {author} {\bibfnamefont {D. S.}~\bibnamefont {Deng}}, \bibinfo {author}
  {\bibfnamefont {A.}~\bibnamefont {Mani}}, \ and\ \bibinfo {author}
  {\bibfnamefont {M.~Z.}\ \bibnamefont {Bazant}},\ }\href@noop {} {\bibfield
  {journal} {\bibinfo  {journal} {Physical Review Letters}\ }\textbf {\bibinfo
  {volume} {107}},\ \bibinfo {pages} {118301} (\bibinfo {year}
  {2011})}\BibitemShut {NoStop}%
\bibitem{KimPRL2015} S. Nam, I. Cho, J. Heo, G. Lim, M. Z. Bazant, D. J. Moon, G. Y. Sung, and S. J. Kim, Physical Review Letters  \textbf{114}, 114501 (2015).
\bibitem{nielsen2014} C. P. Nielsen and H. Bruus,  Physical Review E \textbf{90,} 043020 (2014).
\bibitem{yaroshchuk2011} A. Yaroshchuk, E. Zholkovskiy, S. Pogodin and V. Baulin,  Langmuir \textbf{27,} 11710 (2011).
\bibitem [{\citenamefont {Deng}\ \emph {et~al.}(2013)\citenamefont {Deng},
  \citenamefont {Dydek}, \citenamefont {Han}, \citenamefont {Schlumpberger},
  \citenamefont {Mani}, \citenamefont {Zaltzman},\ and\ \citenamefont
  {Bazant}}]{deng2013overlimiting}%
  \BibitemOpen
  \bibfield  {author} {\bibinfo {author} {\bibfnamefont {D. S.}~\bibnamefont
  {Deng}}, \bibinfo {author} {\bibfnamefont {E.~V.}\ \bibnamefont {Dydek}},
  \bibinfo {author} {\bibfnamefont {J.}~\bibnamefont {Han}}, \bibinfo {author}
  {\bibfnamefont {S.}~\bibnamefont {Schlumpberger}}, \bibinfo {author}
  {\bibfnamefont {A.}~\bibnamefont {Mani}}, \bibinfo {author} {\bibfnamefont
  {B.}~\bibnamefont {Zaltzman}}, \ and\ \bibinfo {author} {\bibfnamefont
  {M.~Z.}\ \bibnamefont {Bazant}},\ }\href@noop {} {\bibfield  {journal}
  {\bibinfo  {journal} {Langmuir}\ }\textbf {\bibinfo {volume} {29}},\ \bibinfo
  {pages} {16167} (\bibinfo {year} {2013})}\BibitemShut {NoStop}%
\bibitem [{\citenamefont {Deng}\ \emph {et~al.}(2015)\citenamefont {Deng},
  \citenamefont {Aouad}, \citenamefont {Braff}, \citenamefont {Schlumpberger},
  \citenamefont {Suss},\ and\ \citenamefont {Bazant}}]{deng2015water}%
  \BibitemOpen
  \bibfield  {author} {\bibinfo {author} {\bibfnamefont {D. S.}~\bibnamefont
  {Deng}}, \bibinfo {author} {\bibfnamefont {W.}~\bibnamefont {Aouad}},
  \bibinfo {author} {\bibfnamefont {W.~A.}\ \bibnamefont {Braff}}, \bibinfo
  {author} {\bibfnamefont {S.}~\bibnamefont {Schlumpberger}}, \bibinfo {author}
  {\bibfnamefont {M.}~\bibnamefont {Suss}}, \ and\ \bibinfo {author}
  {\bibfnamefont {M.~Z.}\ \bibnamefont {Bazant}},\ }\href@noop {} {\bibfield
  {journal} {\bibinfo  {journal} {Desalination}\ }\textbf {\bibinfo {volume}
  {357}},\ \bibinfo {pages} {77} (\bibinfo {year} {2015})}\BibitemShut
  {NoStop}%
\bibitem{schlumpberger2015} S. Schlumpberger, N. B. Lu, M. E. Suss, and M. Z. Bazant, Environ. Sci. Technol. Lett. \textbf{2,} 367 (2015).
\bibitem{conforti2020} K. M. Conforti and M. Z. Bazant, AIChE Journal \textbf{66,} e16751 (2020).
\bibitem{alizadeh2019} S. Alizadeh, M. Z. Bazant and A. Mani, Journal of Colloid and Interface Science \textbf{553,} 451 (2019).
\bibitem{RubinsteinEqui2015} I. Rubinstein  and B. Zaltzman, Phys. Rev. Lett. \textbf{114,} 114502 (2015).
\bibitem{RubinsteinPRF2016} R. Abu-Rjal, I. Rubinstein, and B. Zaltzman, Phys. Rev. Fluid \textbf{1,} 023601 (2016).
\bibitem{RubinsteinPF1991} I. Rubinstein, Physics of Fluids  A  \textbf{3,} 2301 (1991).
\bibitem{Zholkovskij1996} E. K. Zholkovskij, M. A. Vorotyntsev, and E. Staude, J. Colloid Interface Sci. \textbf{181,} 28 (1996).
\bibitem [{\citenamefont {Baygents}\ \emph {et~al.}(1998)\citenamefont
    {Baygents}, \ and\ \citenamefont
    {Baldessari}}]{baygents1998electrohydrodynamic}%
    \BibitemOpen
    \bibfield  {author} {\bibinfo {author} {\bibfnamefont {J.~C.}\ \bibnamefont
    {Baygents}}, \ and\ \bibinfo {author} {\bibfnamefont {F.}~\bibnamefont
    {Baldessari}},\ }\href@noop {} {\bibfield  {journal} {\bibinfo  {journal}
    {Physics of Fluids}\ }\textbf {\bibinfo {volume}
    {10}},\ \bibinfo {pages} {301} (\bibinfo {year} {1998})}.\BibitemShut
    {NoStop}%
\bibitem [{\citenamefont {Aleksandrov}\ \emph {et~al.}(2002)\citenamefont
    {Aleksandrov}, \citenamefont {Grigin},\ and\ \citenamefont
    {Davydov}}]{aleksandrov2002numerical}%
    \BibitemOpen
    \bibfield  {author} {\bibinfo {author} {\bibfnamefont {R.~S.}\ \bibnamefont
    {Aleksandrov}}, \bibinfo {author} {\bibfnamefont {A.~P.}~\bibnamefont
    {Grigin}}, \ and\ \bibinfo {author} {\bibfnamefont {A.~D.}~\bibnamefont
    {Davydov}},\ }\href@noop {} {\bibfield  {journal} {\bibinfo  {journal}
    {Russian Journal of Electrochemistry}\ }\textbf {\bibinfo {volume}
    {38}},\ \bibinfo {pages} {1097} (\bibinfo {year} {2002})}.\BibitemShut
    {NoStop}%
\bibitem [{\citenamefont {Storey}\ \emph {et~al.}(2007)\citenamefont
   {Storey}, \citenamefont {Zaltzman},\ and\ \citenamefont
   {Rubinstein}}]{story2007bulk}%
   \BibitemOpen
   \bibfield  {author} {\bibinfo {author} {\bibfnamefont {B.~D.}\ \bibnamefont
   {Storey}}, \bibinfo {author} {\bibfnamefont {B.}~\bibnamefont
   {Zaltzman}}, \ and\ \bibinfo {author} {\bibfnamefont {I.}~\bibnamefont
   {Rubinstein}},\ }\href@noop {} {\bibfield  {journal} {\bibinfo  {journal}
   {Physical Review E}\ }\textbf {\bibinfo {volume}
   {76}},\ \bibinfo {pages} {041501} (\bibinfo {year} {2007})}.\BibitemShut
   {NoStop}
\bibitem{bazant2004} M. Z. Bazant, Proc. Roy. Soc. {\bf 460,} 1433  (2004).
\bibitem{Gu2019} Z. B. Gu, B. R. Xu, P. Huo, S. Rubisntein, M. Z. Bazant, and D. S. Deng. Phys. Rev. Fluid \textbf{4,} 113701 (2019).	
\bibitem{bazant2003} M. Z. Bazant, J. Choi, and B. Davidovitch, Phys. Rev. Lett. \textbf{91,} 045503 (2003).
\bibitem{Maniphyfluid2013}  C. Druzgalski, M. B. Andersen, and A. Mani, Physics of Fluids \textbf{25,} 8316 (2013).
\bibitem{ManiJCIS2015} E. Karatay, C. L. Druzgalski, and A. Mani, Journal of Colloid and Interface Science \textbf{446,} 67 (2015).
\bibitem{LiuJAP2018} P. Shi and W. Liu, Journal of Applied Physics  \textbf{124,} 204304 (2018).
\bibitem{LiuPRE2020} W. Liu, Y. Zhou, and P. Shi, Phys. Rev. E \textbf{101,} 043105 (2020).
\bibitem{Greenleea2009Reverse}L. F. Greenleea, D. F. Lawlerb, B. D. Freemana, B. Marrotc, and P. Moulinc, Water Research \textbf{43,} 2317 (2009).
\bibitem{Achilli2009Power}A. Achilli, T. Y. Cathb, and A. E. Childressa, Journal of Membrane Science \textbf{343,} 42 (2009).
\end{thebibliography}
\end{document}